\documentclass[sigconf]{acmart}

\renewcommand\footnotetextcopyrightpermission[1]{}
\acmConference[FSE'26 Industry]{2026 ACM International Conference on the Foundations of Software Engineering (FSE)}{July 5--9, 2026}{Montreal, Canada}

\newcommand{\toolName}{\texttt{BayesInsights}}

\begin{document}

\title{BayesInsights: Modelling Software Delivery and Developer Experience with Bayesian Networks at Bloomberg}

\author{Serkan Kirbas}
\affiliation{%
  \institution{Bloomberg}
  \city{London}
  \country{United Kingdom}}
\email{skirbas@bloomberg.net}

\author{Federica Sarro}
\affiliation{%
  \institution{UCL, Bloomberg}
  \city{London}
  \country{United Kingdom}}
\email{f.sarro@ucl.ac.uk}

\author{David Williams}
\affiliation{%
  \institution{UCL, Bloomberg}
  \city{London}
  \country{United Kingdom}}
\email{david.williams.22@ucl.ac.uk}

\begin{abstract}
As software in industry grows in size and complexity, so does the volume of engineering data that companies generate and use. Ideally, this data could be used for many purposes, including informing decisions on engineering priorities. However, without a structured representation of the links between different aspects of software development, companies can struggle to identify the root causes of deficiencies or anticipate the effects of changes.

In this paper, we report on our experience at Bloomberg in developing a novel tool, dubbed \toolName, which provides an interactive interface for visualising causal dependencies across various aspects of the software engineering (SE) process using Bayesian Networks (BNs). We describe our journey from defining network structures using a combination of established literature, expert insight, and structure learning algorithms, to integrating \toolName{} into existing data analytics solutions, and conclude with a mixed-methods evaluation of performance benchmarking and survey responses from 24 senior practitioners at Bloomberg.

Our results revealed 95.8\% of participants found the tool useful for identifying software delivery challenges at the team and organisational levels, cementing its value as a proof of concept for modelling software delivery and developer experience. \toolName{} is currently in preview, with access granted to seven engineering teams and a wider deployment roadmap in place for the future.
\end{abstract}



\maketitle

\section{Introduction} \label{sec:intro}
At Bloomberg, the DevX team distributes quarterly surveys to software engineers regarding the state of the firm's software engineering (SE) practice. The survey comprises Likert-scale questions capturing engineers’ perceptions of their work environment, including satisfaction and engagement, and covers topics such as how meaningful they find their work, how much time is lost due to obstacles, the quality of peer feedback, as well as aspects of the SPACE framework~\cite{forsgrenSPACE} and the DevOps Research and Assessment (DORA) metrics~\cite{forsgren-2018-accelerate, DORAReport2024}. The results of these surveys, along with contribution metrics, are visualised via dashboards and reporting tools. While such dashboards provide visibility into delivery performance, they do not explain why certain outcomes occur or how changes in one area might affect others. For example, a change in one metric, such as reduced deployment frequency, may reflect improvements in related factors like CI/CD or automated testing, making interpretation ambiguous~\cite{forsgren-2018-accelerate}. As a result, teams can struggle to estimate trade-offs or predict the impact of engineering changes.

In response to this, we investigated the use of Bayesian Networks (BNs) for causal analysis in SE practices. We propose an interactive tool, dubbed \toolName, to model causal and probabilistic dependencies among key SE factors at Bloomberg. We use data from one internal engineering survey (20 questions, >2,000 responses) as the source for our BNs, in which each question is represented as a graph node (with the node states corresponding to the question options). We derive the structure of links between these nodes using a hybrid and iterative modelling approach consisting of \textcircled{1} reviewing established literature (\S\ref{sec:structure_lit}), \textcircled{2} refinement through expert opinion (\S\ref{sec:structure_expert}), and \textcircled{3} validation using structure learning algorithms (\S\ref{sec:structure_quant}).

\begin{figure*}[t]
\centering
\includegraphics[width=0.85\textwidth]{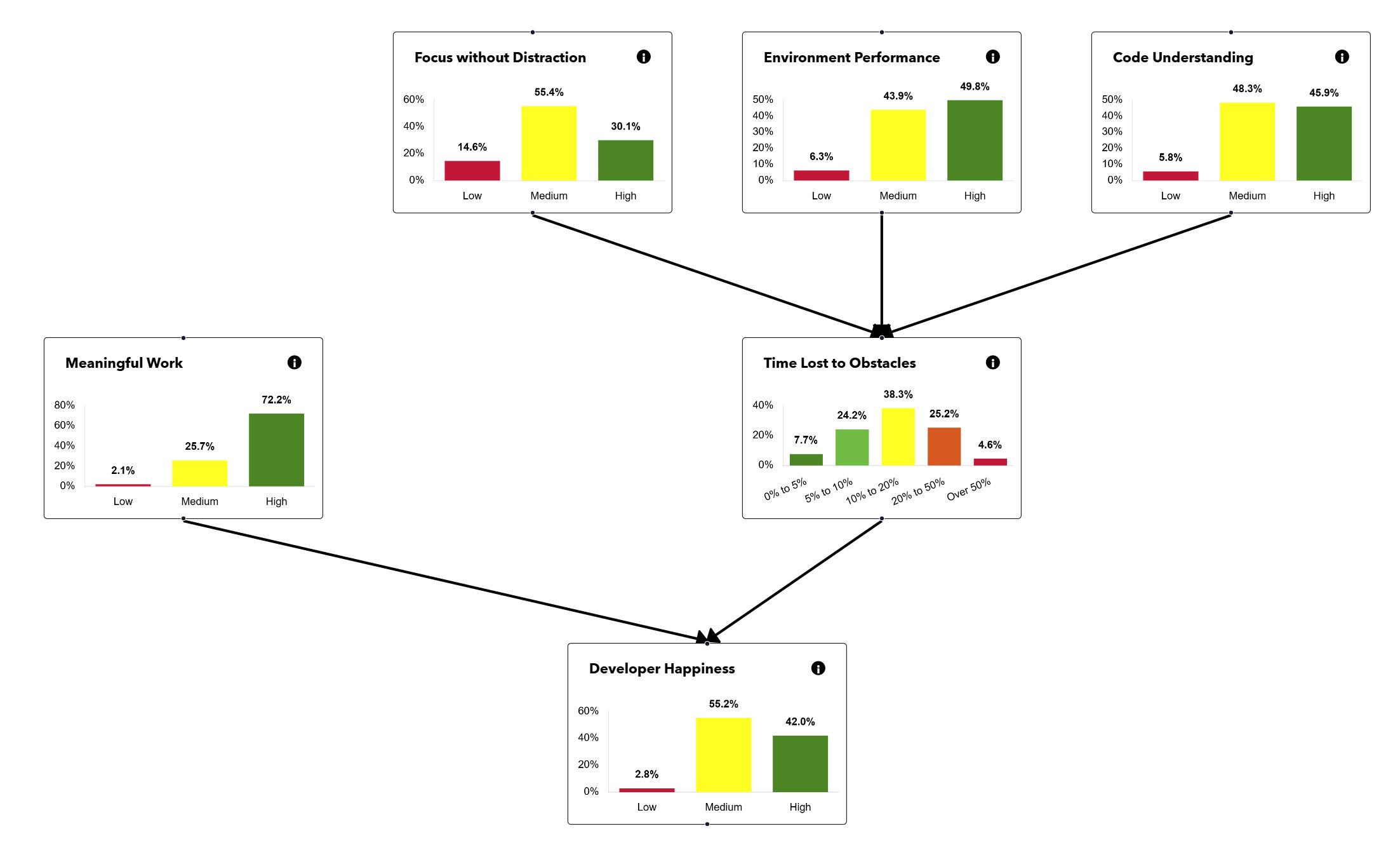}
  \caption{The DevEx Bayesian Network in \toolName{} showing causal links among factors influencing developer happiness. Users can explore “what-if” scenarios by selecting a bar within a node, fixing that outcome and updating the network to reflect its impact on related factors. Mocked probability distributions are shown to protect sensitive data; values are illustrative only.}
  \Description{The DevEx Bayesian Network in \toolName{} showing causal links among factors influencing developer happiness. Users can explore “what-if” scenarios by selecting a bar within a node, fixing that outcome and updating the network to reflect its impact on related factors. Mocked probability distributions are shown to protect sensitive data; values are illustrative only.}
  \label{fig:bayesiannetwork}
\end{figure*}

Figure~\ref{fig:bayesiannetwork} illustrates an example of a BN in \toolName{} (with mocked probability distributions to protect sensitive data). Here, we can see that factors such as \textit{focus without distraction}, \textit{environment performance}, and \textit{code understanding} influence \textit{time lost to obstacles}, while both \textit{time lost to obstacles} and \textit{meaningful work} contribute directly to \textit{developer happiness}. With \toolName, users can conduct impact analyses through “what-if” scenarios by selecting evidence. For instance, users can estimate the extent to which minimising time lost to obstacles would improve developer happiness, or how maximising development environment performance cascades through the system to influence outcomes.

We conducted a mixed-methods evaluation of \toolName{} that combines quantitative performance benchmarking with qualitative feedback collected through a post-session survey following focus group sessions with 28 senior Bloomberg practitioners (24 of whom responded). \toolName{} demonstrated both practical responsiveness and strong perceived usefulness: 95.8\% reported that \toolName{} helps identify delivery challenges at the team or organisational level, and 75\% found the outputs easy to interpret. Moreover, 79.2\% would use or recommend the tool, and many suggested avenues for improvement, motivating further development. \toolName{} is currently deployed in early access to seven engineering teams, with a development roadmap in place for wider deployment in future. This paper provides an in-depth description of our BN creation methodology and insights into the value of causal modelling for SE practice at a large-scale software company.

\section{Bayesian Network Structure Definition} \label{sec:structure}
For the initial implementation of \toolName, we developed two BNs based on internal engineer survey data. The first BN models \textbf{software delivery performance}, represented by two dimensions: \textbf{throughput}, which captures the speed and frequency of delivering changes, and \textbf{stability}, which reflects the reliability of changes. Throughput is modelled using \textit{change lead time}, \textit{deployment frequency}, and \textit{failed deployment recovery time}, while stability is represented by \textit{change failure rate}. These metrics form the core model and are influenced by other upstream engineering practices. The second BN modelling \textbf{DevEx (Developer Experience)}\footnote{In this context, DevEx refers to the satisfaction and engagement of engineers within their work environment.} and the factors that shape it (Figure~\ref{fig:bayesiannetwork}) is derived from other aspects covered in the internal engineering survey, such as perceptions of \textit{meaningful work}, \textit{time lost to obstacles}, and \textit{focus without distraction}.

Constructing Directed Acyclic Graphs (DAGs) for BNs in SE contexts presents significant challenges~\cite{Siebert_StatsCausal}. Software delivery metrics exhibit complex interdependencies that are rarely observed in isolation. Hidden confounders, including organisational culture, toolchain maturity, and team composition, can create spurious associations between variables. This challenge is compounded by the inherent limitations of SE data. Surveys are prone to interpretation and recall bias, and the temporal nature of software development processes further complicates causal inference, as relationships between variables may differ across teams and evolve with project or organisational changes. These factors collectively render naive, data-only structure learning approaches unreliable and highlight the necessity of a multi-source methodology to construct credible DAG structures for causal inference~\cite{Siebert_StatsCausal, Daly2011LearningBN}.

To devise our network structures, we adopted a hybrid modelling approach combining \textit{theoretical insights from the literature}, \textit{expert-defined causal links}, and \textit{algorithmic learning techniques}, following prior work on integrating expert input with data-driven methods~\cite{misirliMappingStudyOnBN}. This process involved an iterative cycle of structure proposal, expert validation, and refinement. With this approach, we aim to achieve a comprehensive, context-aware representation of causal dependencies that goes beyond what standard DevOps metric dashboards provide. We detail each step below.

\subsection{Established Literature} \label{sec:structure_lit}
We began with a foundational DAG based on the DORA metrics~\cite{DORAReport2024} and their established causal links. Additional nodes were created from factors measured in one of Bloomberg's quarterly internal engineering surveys, which captures engineering outcomes, processes, and DevEx. Standard definitions of factors covered in this study are provided in our online repository~\cite{repo}. Each node in the BN maps to a specific survey question, meaning every node represents a measurable aspect of engineering practice. For example, the question \textit{``Technical debt did not significantly impact my ability to complete new work''} was mapped to the node \textit{`Tech Debt Impact'}. We hypothesised causal links between nodes using a process adapted from Pearl's causal hierarchy and established BN methods~\cite{fentonRiskAssessmentBN,Pearl2000}. We used the following causal modelling idioms~\cite{neil2000building}: \textbf{\textcircled{1} Cause-Consequence idiom}, where one variable plausibly influences another through a physical, productive, or intentional mechanism; and \textbf{\textcircled{2} Definition/Synthesis idiom}, where a concept is directly determined by, or made up of, its components. We defined a causal hypothesis for each proposed link in the DAG either using the idioms or through prior literature, and then evaluated them using expert opinion and algorithmic learning techniques described in the following sections.

\subsection{Expert Insight Survey} \label{sec:structure_expert}
After developing our node-link hypotheses based on the established literature, we sought expert insight to refine our notions and account for Bloomberg-specific practices. To achieve this, we conducted a structured survey that covered directional cause-and-effect relationships among the factors measured in the internal engineer survey. Each question, representing a possible link between two factors in the causal graph, asked whether one factor directly influences the other in practice. For example, a question assessing the link \textit{Time Lost to Obstacles $\rightarrow$ Developer Happiness} was phrased as ``To what extent does time lost to obstacles influence developer happiness?'' All questions featured the same set of options for level of influence: ``Strong'', ``Moderate'', ``Weak'', ``None'', and ``Not sure/No opinion''. In total, our survey covered 24 potential relationships and allowed experts to suggest any additional ones.

Participants were purposely selected from Bloomberg’s DevX team ($n=8$) for their domain expertise and organisation-wide contextual understanding, with an average of 10 years of experience at Bloomberg. As part of their roles, they regularly work with engineering metrics and development workflows, making them well-suited to assess the relationships in question. While the sample size is small, this is typical in expert surveys where the focus is on depth of insight rather than broad statistical generalisation~\cite{gaagElicitProb}.

After collecting responses, we calculated a weighted score for each proposed relationship based on participants' influence ratings. Responses marked as ``Strong'' were given a weight of 1.0, ``Moderate'' as 0.8, ``Weak'' as 0.2, and all other responses (``None'' or ``Not sure'') as 0. These values were averaged across all responses to produce a final score for each relationship, and after further consultation with participants, we set 0.70 as the minimum score for a relationship to be retained in the final BNs. Based on the scoring results, two relationships were removed, and three were introduced, yielding the final structures of the expert-refined BNs. To protect sensitive expert knowledge and internal development practices, the survey questions and results are omitted from this study.

\subsection{Structure Learning Algorithms} \label{sec:structure_quant}
To further refine the DAG structures, we applied the  Hill-Climbing (HC) and Peter-Clark (PC) structure-learning algorithms (described in \S\ref{sec:bg_structure}) to the internal engineering survey data to empirically estimate the BN structures. These methods detect statistical dependencies in the data and surface additional candidate edges not previously suggested by theory or expert input. To account for sampling variability, we used bootstrap resampling, repeatedly generating new samples from the data and assigning a confidence score to each recovered edge. This provided a stability check, highlighting consistently supported links versus those likely due to noise. We evaluated the resulting structures by computing Bayesian Information Criterion (BIC) scores, which reward goodness of fit while penalising unnecessary complexity (with lower values indicating better model fit). We compared the BIC scores for the HC and PC structures to those of the expert-refined models and found that the expert models performed best, followed by HC, and finally PC. HC showed partial overlap with the expert structures and was most informative when constrained by DORA definitions, suggesting potential refinements worth expert consideration, whereas PC added little value due to weak overlap and a poorer BIC. This pattern is consistent with studies showing that constraint-based learners degrade quickly with noisier data and that adding hard constraints improves HC accuracy~\cite{SCUTARI2019235}. Finally, considering that scoring algorithms cannot be relied upon to guarantee causal validity~\cite{Kitson2023Survey}, the expert-refined structures, which also achieved the best BIC scores, remained the most reliable foundation, with HC outputs used only as edge candidates for refinement subject to expert review.

\subsection{Finalised Network Structure \& Conditional Probability Table Computation} \label{sec:structure_cpt}
Not all sources we explored were given equal weight in determining the final network structures. We prioritised the DORA report~\cite{DORAReport2024} and expert opinions when deciding which relationships to include and how confident we were in them. The strongest relationships are those supported by multiple high-quality sources. By starting with a solid foundation based on DORA and carefully integrating established literature, expert feedback, and algorithmic results, our models are both grounded in theory and responsive to real-world data, while avoiding overreliance on any single source.

After defining the final model structures, each node was parameterised by computing Conditional Probability Tables (CPTs) to capture the dependencies between nodes. To estimate these probability distributions, we used the internal engineering survey data (20 questions, >2,000 responses). Since the data stem from Likert-scale questions, we treat them as discrete categorical variables. For root nodes (those without parents), probabilities were estimated as the marginal distribution: $P(X = x) = \frac{\text{count}(X = x)}{N}$, where $N$ is the total number of valid observations. For child nodes, conditional probabilities were estimated from co-occurrence counts with parent configurations, e.g., a node with two parents $A$ and $B$: $P(X = x \mid A = a, B = b) = \frac{\text{count}(X = x, A = a, B = b)}{\text{count}(A = a, B = b)}$. To address sparsity in rare or unobserved parent-child combinations, we applied Bayesian-Dirichlet equivalent uniform (BDeu) smoothing, preventing zero-probability CPT entries. Here, $P=p$ denotes a specific joint configuration of all parent nodes of node $X$: $P(X = x \mid \text{Parents} = p) = \frac{\text{count}(X = x, P = p) + \alpha}{\text{count}(P = p) + \alpha K}$, where $\alpha$ is the equivalent sample size controlling the strength of the Dirichlet prior, and K is the number of discrete states of the child node.

\section{Implementation \& Integration} \label{sec:design}
After finalising the structures and CPTs for both the software delivery performance and DevEx BNs, our goal was to present them in an intuitive way so early-access users within the DevX team could experience them. We used a fully open source tech stack: \toolName{} is implemented as a client-server architecture built with \texttt{Django}~\cite{tech-django} and \texttt{Django Ninja}~\cite{tech-django-ninja}, exposing RESTful endpoints which return BN structures and process evidence-based inference queries. The backend retrieves and preprocesses the internal engineering survey data before model training and inference using \texttt{pgmpy}~\cite{tech-pgmpy,ankanPgmpy}. The front-end for visualising and interacting with the BNs is implemented in TypeScript. Networks are rendered using React Flow~\cite{tech-reactflow}, with nodes displayed as Chart.js~\cite{tech-chartjs} bar charts showing probability distributions. This UI was integrated alongside existing metric dashboards, inheriting robust access control policies and complementing other software analytics solutions.

We designed \toolName{} as an interactive tool. It presents a dynamic, directed graph in which key metrics are represented as connected nodes, forming a BN. It visualises how different engineering metrics influence one another based on historical survey data, enabling users to conduct powerful ``what-if?'' scenario analysis to identify the most valuable areas to target to improve delivery performance/engineer satisfaction. For example, with the software delivery performance BN, managers can investigate how a change in lead time might affect their team's delivery throughput. Each node shows the probability distribution of outcomes and updates in real time as users select evidence, triggering a back-end query and re-rendering the network with updated probabilities.

\section{Evaluation} \label{sec:eval}
To evaluate \toolName{}, we conducted quantitative performance testing to assess its practical usability and gathered qualitative insights from focus groups with prospective users to establish its value in providing insights into software delivery practice.

\subsection{Performance Testing}\label{sec:eval_quant}
To assess the responsiveness of the deployed \toolName{} service, we conducted performance testing using the Hyperfine~\cite{tech-hyperfine} benchmarking tool for single-request latency and Locust~\cite{tech-locust} for concurrent load testing. For single inference requests, the average response time was 24 ms. Under load with 50 concurrent users, median response times were under 40 ms. These results confirm that the tool is suitable for real-time exploration of BNs.

\subsection{User Study}\label{sec:eval_qual}
\subsubsection{Methodology}\label{sec:eval_qual_methodology}
To understand the value of \toolName, we conducted a series of focus group sessions. Each 45-minute session began with a short presentation familiarising participants with the tool, including the challenge it addresses, the modelling method, and its potential for reasoning about delivery performance. This was followed by a live demo during which participants were introduced to the interface and taught how to enter evidence and interpret the outputs. Participants were then given access to \toolName{} and encouraged to explore scenarios on their own. Finally, feedback was collected through a short, anonymous questionnaire comprising 22 questions: 4 related to participants' roles, development experience, and familiarity with software delivery and DevEx metrics, and 18 eliciting their opinions on the tool in terms of interpretability, model quality and trust, usefulness and practical application, overall feedback, and suggestions for improvements. The full set of questions is provided in our repo~\cite{repo}.

\subsubsection{Results}\label{sec:eval_qual_results}
We conducted seven sessions, each with 3--5 participants. In total, 28 attended the sessions, and 24 completed the questionnaire. Given the potential managerial value of \toolName{}, we primarily targeted senior practitioners: Of those who completed the questionnaire, 21 had more than 10 years of experience, while the remaining three had 4--10 years. This cohort spanned 8 Team Leads, 6 Software Managers, 5 Product Managers, 4 Software Engineers (including one DevX Engineer), and 1 Manager of Coaching.

Responses indicated that participants found \toolName{} accessible and easy to understand. Most (75\%) stated that the outputs were easy to interpret, and a large majority (83.3\%) clearly understood how the model visualised changes, particularly how altering one metric affected others in what-if analyses. Participants valued how the tool clearly visualised relationships between metrics and highlighted its dynamic nature, with one praising the ``ability to change inputs and get immediate feedback'' and another remarking that ``seeing the most significant impacts [helps] with prioritising improvements.'' This functionality was regarded as particularly useful because it was ``grounded in real data'' and aligned with practical ways of reasoning about delivery challenges. Trust in the tool was also positive, with 70.9\% of respondents expressing confidence in its outputs when reasoning about software delivery and DevEx.

Almost all respondents (95.8\%) reported that it was useful for understanding delivery challenges at a team or organisational level. Many mentioned contexts in which they would use \toolName, such as reviewing team practices (37.5\%), supporting leadership decisions (62.5\%), identifying root causes of delivery issues (50\%), and running what-if scenarios (62.5\%). A strong majority (79.2\%) also indicated that they would use or recommend \toolName{} in their own teams. Notably, 25\% of participants reported that even within the 15-minute demonstration period, the tool helped them generate concrete ideas for improving delivery performance (e.g., ``advocating for fewer distractions and more release ease'', ``exploring ways to enable focus and deep work with [my] team''). This highlights immediate value and suggests that prolonged use could help teams identify additional avenues to improve their practices. 

Participants also identified a few areas for improvement. The most frequent suggestion was to incorporate natural language summaries of the tool's outputs, offering clearer explanations of what had changed and why, along with possible actions informed by internal documentation on best practices. Another recurring request was for a results-first interface layout, in which inputs and outputs are shown above, with the BN diagram presented below as a reference. Several participants expressed interest in having an indication of the scale of and confidence in the influence between factors, and in being able to view their team's data alongside the overall network for comparison. Overall, participants' suggestions highlight the need to foreground explainability, transparency, and practicality when using causal techniques in a user-facing tool.

\section{Journey \& Outcomes} \label{sec:journey}
\toolName{} is the result of a joint effort between Bloomberg and University College London (UCL), a long-standing partnership that has led to several cutting-edge industrial projects and publications~\cite{williams2024bloomberg}. The project was completed over the summer of 2025 with participation from the DevX Insights team at Bloomberg, as well as one professor, one Ph.D. student and six master's students from UCL Computer Science via the Industry Exchange Network (IXN) programme~\cite{ucl-ixn}. UCL members were onboarded as SE contractors at Bloomberg to contribute to the design and development of \toolName{} and to conduct the user studies to evaluate it. As of October 1, 2025, \toolName{} was made available in early access to seven teams within DevX, and development roadmaps have been put in place to continue its development towards a state suitable for wider deployment across the organisation. Although this implementation is in the early stages, feedback received throughout the project has been highly encouraging, with keen interest from senior engineers, team leads, and managers. We posit that a major component of this project's success is the nature of industrial-academic collaboration, which enabled academics to explore and evaluate techniques not yet deemed industry-standard in a practical setting using large-scale engineering data. Through this partnership, we demonstrated the value of causal techniques for modelling SE delivery and developer experience in a real-world setting.

\section{Related Work} \label{sec:bg} 
This section discusses SE metrics, outlines challenges in interpretation, and reviews causal modelling approaches, focusing on BNs, their alternatives, and structure learning in SE modelling.

\textit{\textbf{Interpreting Software Metrics.}} \label{sec:bg_difficulties}
Software delivery performance is often benchmarked with DORA metrics~\cite{DORAReport2024}, along with measures such as test coverage, code review speed, and technical debt management~\cite{DORAReport2024,wang2022testautomationmaturityimproves}. DevEx metrics complement delivery outcomes by capturing the human side of engineering work, with frameworks like SPACE covering satisfaction, performance, and collaboration~\cite{DORAReport2024,forsgrenSPACE,meyerDailyLifeofSDE}. These metrics provide valuable benchmarks, but they are usually tracked separately. Delivery data often comes from pipelines and incidents~\cite{hughes_2021_cicd}, while DevEx relies on surveys or HR systems, which makes them difficult to integrate.

Software teams are also complex socio-technical systems shaped by many confounders. As noted in Goodhart's Law~\cite{Goodhart1984ProblemsOM}, focusing on individual indicators can be misleading. In addition, the lack of formal causal reasoning limits the ability to perform systematic, data-driven analysis and hinders the exploration of how specific changes may influence engineering outcomes~\cite{ieeeCausality2023}. This gap highlights the potential of BNs, which can model dependencies explicitly and enable data-driven reasoning under uncertainty.

\textit{\textbf{Bayesian Networks in Software Engineering.}} \label{sec:bg_bn_in_se}
BNs model probabilistic dependencies between variables in a DAG and support forward and backward causal reasoning~\cite{KollerCPT, heckermanTutorialBN}, making them powerful for root-cause analysis and scenario exploration under uncertainty~\cite{fentonRiskAssessmentBN}. Rather than evaluating metrics in isolation, BNs represent interacting influences and propagate their effects throughout a system. In SE contexts, BNs have been applied to defect prediction, effort estimation, and project management, but these have typically been narrow in scope and challenging to generalise across teams or projects~\cite{deSousa2022mapping, icsInsightsOfBN}. Little work has addressed how BNs can unify delivery metrics with human-centred outcomes~\cite{deSousa2022mapping, misirliMappingStudyOnBN}. Given evidence that delivery and DevEx are coupled, BNs are promising for modelling trade-offs and dependencies across both domains.

\textit{\textbf{Alternative Techniques for Causal Reasoning.}} \label{sec:bg_alternatives}
Researchers have used statistical causal inference (SCI) methods for causal reasoning in SE, such as propensity score matching (PSM) and difference-in-differences (DiD). A recent mapping study found only 25 SE papers (published between 2010 and 2022) using any SCI method, with fewer than 10 applying the classical designs mentioned above~\cite{Siebert_StatsCausal}. Most of them focused on code quality or defects, with little attention to human-centred outcomes like developer satisfaction. Another review of 45 causal studies in SE found that two used DiD, one used PSM, and five used BN models~\cite{ieeeCausality2023}. While valuable, DiD and PSM rely on assumptions (i.e., that all relevant group influences are measured, and that groups affected and unaffected by an intervention would have otherwise followed the same trajectory, respectively) that often do not hold in SE contexts ~\cite{PearlGlymourJewell2016CausalPrimer,Imbens_Rubin_2015_Causal_Inference_Stats, Angrist_DiD}.

\textit{\textbf{Structural Learning Algorithms.}} \label{sec:bg_structure}
Building appropriate BN structures poses a significant challenge. Prior reviews on BN applications in SE have highlighted that most structures rely solely on expert knowledge, with few using data-driven approaches~\cite{misirliMappingStudyOnBN, tosunSystematicLitOnBN}. Such reliance limits reproducibility and risks embedding bias. While data-driven learning offers an alternative, SE survey data can be noisy and biased, making reliable structure recovery difficult~\cite{molleri2020empirically}. Thus, some studies advise constraining algorithms with expert input and carefully verifying results~\cite{Kitson2023Survey}. Two main families of algorithms exist. Constraint-based methods, such as the Peter-Clark (PC) algorithm, use independence tests to remove unsupported edges, producing graphs that avoid many false links but risk missing true ones~\cite{Kitson2023Survey}. In contrast, score-based methods, such as the Hill Climbing (HC) algorithm, search for graphs balancing fit and simplicity~\cite{math11112524}. Although these methods capture more relationships, they risk overfitting data and including spurious edges. Hybrid algorithms combine the two, using constraint tests to narrow candidates before applying scoring to refine them~\cite{Kitson2023Survey}. In practice, researchers rely on (i) how well the network explains new data, often quantified using the Bayesian Information Criterion (BIC), which balances model fit against structural complexity to avoid overfitting~\cite{Kitson2023Survey}, (ii) stability checks such as bootstrapping to see which edges consistently reappear~\cite{FriedmanBootstrap}, and (iii) expert review of whether inferred links make sense~\cite{misirliMappingStudyOnBN, BNnoisydata}.

In this work, we first construct a hypothesised DAG informed by literature, and refine it with expert input. For further refinement, we apply structure learning (PC and HC) with bootstrapping to assess edge stability, and compare models using BIC scoring.

\section{Conclusion} \label{sec:conclusions}
This study presents \toolName, an interactive tool developed and released in preview at Bloomberg that uses Bayesian Networks to allow users to explore causal relationships across various aspects of the software engineering process. Our testing and prospective user evaluations demonstrate the tool's practical viability and effectiveness in revealing bottlenecks and high-impact areas of improvement in software engineering performance and developer satisfaction.

\section*{Data Availability}
As the BN has been developed within Bloomberg’s internal systems and repositories, data and source code are not publicly accessible. However, the technologies we used in the project are all open source, and we explained in detail how to implement such a tool in practice, so that future academics and practitioners alike can adopt it. Moreover, we make available the questionnaire we distributed to carry out the user evaluation~\cite{repo}.

\section*{Acknowledgments}
We thank all Bloomberg engineers who participated in our user studies and the UCL alumni \textit{Shahzeb Ahmad}, \textit{Mizbah Celik}, \textit{Sultan Insan Geosrinov Muhammad Rifqi}, \textit{Narmin Mustafali}, \textit{Ting-Yun Wang}, and \textit{Farhan Zaki},
who contributed to this project at Bloomberg through the UCL Industry Exchange Network (IXN). 

This work received ethics approval from Bloomberg and University College London (LREC no.: 2025-1647).


\bibliographystyle{ACM-Reference-Format}
\bibliography{Bibliography}
 
\end{document}